\documentclass[conference]{IEEEtran}
\IEEEoverridecommandlockouts
\usepackage{textcomp}
\usepackage{xcolor}
\usepackage{cite}
\usepackage{latexsym}
\usepackage{graphicx}
\usepackage{amsfonts,amssymb,amsmath}
\usepackage{nccmath}
\usepackage{optidef}
\usepackage{hyperref}
\usepackage{color,soul}
\usepackage{array}
\usepackage{algorithm}
\usepackage{algorithmic}
\usepackage[T1]{fontenc}
\usepackage[utf8]{inputenc}
\usepackage{fancyhdr}
\usepackage{lastpage}
\usepackage{caption}
\usepackage{subcaption}

\DeclareMathAlphabet{\pazocal}{OMS}{zplm}{m}{n}

\newcommand{\Ab}{\pazocal{A}}
\newcommand{\Mb}{\pazocal{M}}

\newcommand{\Bb}{\pazocal{B}}

\newcommand{\Ib}{\pazocal{I}}

\def\BibTeX{{\rm B\kern-.05em{\sc i\kern-.025em b}\kern-.08em
    T\kern-.1667em\lower.7ex\hbox{E}\kern-.125emX}}
\begin{document}

\title{Enhanced User Grouping and Pairing Scheme for CoMP--NOMA-based Cellular Networks\\
}

\author{
    \IEEEauthorblockN{Akhileswar Chowdary \IEEEauthorrefmark{1}, Garima Chopra \IEEEauthorrefmark{1}, Abhinav Kumar\IEEEauthorrefmark{1}, and Linga Reddy Cenkeramaddi\IEEEauthorrefmark{2}}
    \IEEEauthorblockA{\IEEEauthorrefmark{1} Department of Electrical Engineering, Indian Institute of Technology Hyderabad, Telangana, 502285 India.
    \\Email: ee19mtech11028@iith.ac.in, \{garima.chopra, abhinavkumar\}@ee.iith.ac.in}
    \IEEEauthorblockA{\IEEEauthorrefmark{2}Department of Information and Communication Technology, University of Agder, Grimstad, 4879 Norway.
    \\Email: linga.cenkeramaddi@uia.no}
}
\maketitle
\begin{abstract}
Non-orthogonal multiple access (NOMA) has been identified as one of the promising technologies to enhance the spectral efficiency and throughput for the fifth generation (5G) and beyond 5G cellular networks. Alternatively, Coordinated multi-point transmission and reception (CoMP) improves the cell edge users' coverage. Thus, CoMP and NOMA can be used together to improve the overall coverage and throughput of the users. However, user grouping and pairing for CoMP--NOMA-based cellular networks have not been suitably studied in the existing literature. Motivated by this, we propose a user grouping and pairing scheme for a CoMP--NOMA-based system. Detailed numerical results are presented comparing the proposed scheme with the purely OMA-based benchmark system, NOMA only, and CoMP only systems. We show through simulation results that the proposed scheme offers a trade-off between throughput and coverage as compared to the existing NOMA or CoMP based system.\\
\end{abstract}


\section{Introduction}
Non-orthogonal multiple access (NOMA) has emerged as the promising multiple access scheme to enhance the spectral efficiency (in turn the network throughput) for the 5G and beyond 5G cellular networks. The primary idea of NOMA is to serve multiple users utilizing the same resource (e.g. spectrum, time, etc.) at the cost of increased complexity due to successive interference cancellation (SIC) \cite{n1}, \cite{n2}. In a typical downlink power-domain NOMA, the Base Station (BS) sends the superpositioned signals of the users which differ in their power. The user decodes its intended message either by treating information of another user sharing the same resource as noise or through SIC.

Coordinated multi-point with joint transmission and reception which hereafter is referred to as CoMP has been extensively researched in today's wireless networks \cite{c1}. It has been shown in \cite{c2}, that joint transmission CoMP improves the network coverage at the cost of reduced network throughput. This throughput reduction in CoMP can be mitigated by utilizing NOMA with CoMP networks \cite{c3}. Using joint transmission NOMA (JT-NOMA) for the CoMP system, in this paper, we analyze NOMA for CoMP as well as non-CoMP users to improve the network coverage and throughput.

\begin{figure}[t]
    \centering
    \includegraphics[width=9cm,height=12cm,keepaspectratio]{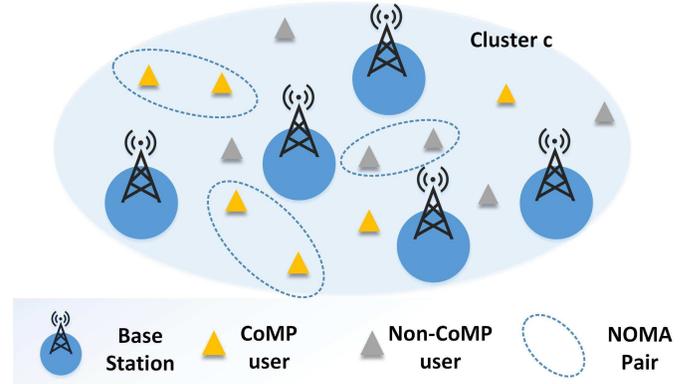}
    \caption{System Model}\vspace{-0.2in}
    \label{fig:system_model}
\end{figure}
The existing works on CoMP--NOMA system in \cite{l4,l6,l1,l3,l5,l7} have not studied the performance of the network for different types of user pairs possible, namely, CoMP--CoMP, (non-CoMP)--CoMP, (non-CoMP)--(non-CoMP), for a generalized scenario (considering multiple cells, randomly distributed users, and BSs). In a cluster of BSs, after differentiating CoMP and non-CoMP users, grouping the users for performing NOMA pairing is not obvious. As shown in \cite{aaa}, a CoMP user cannot act as both strong and weak user when paired with multiple non-CoMP users. Thus, considering such conditions, the grouping of CoMP and non-CoMP users before NOMA pairing is a non-trivial task. Furthermore, after forming the aforementioned hybrid pairs of users, scheduling them is also not straightforward. Motivated by this, we perform a detailed study of throughput and coverage of CoMP--NOMA-based systems for a grouping and pairing scheme. The main contributions of the paper are as follows:
\begin{enumerate}
    \item We propose a user grouping and pairing scheme to study the performance of CoMP--CoMP and (non-CoMP)--(non-CoMP) NOMA pairs considering network scaling.
    \item After pairing users from the groups formed, we perform scheduling for a CoMP--NOMA-based system.
    \item The performance of the proposed scheme is compared with the purely OMA-based benchmark system, CoMP only, and NOMA only systems. We show that the proposed scheme results in superior throughput than the CoMP only system and superior coverage than the NOMA only system.
\end{enumerate}

The organization of the paper is as follows. Section \ref{system_model} describes the system model in detail. Section \ref{pairing} presents the proposed user grouping and pairing scheme for CoMP--NOMA-based system. The simulation results are discussed in Section \ref{results}. Finally, the work is concluded in Section \ref{conclusion}.
\section{System Model} \label{system_model}
We consider a downlink cellular network where BSs and users are randomly deployed according to a homogeneous Poisson point process (PPP) with densities $\lambda_b$ and $\lambda_u$, respectively \cite{ppp} as shown in Fig. \ref{fig:system_model}. Let $\mathcal{B}=\lbrace 1,2,...,B \rbrace$ be the set of BSs deployed in an area $\mathcal{A}$. Let $\mathcal{M}=\lbrace 1,2,...,\text{M} \rbrace$ be the set of subchannels for a given BS $b$. The users are associated with the BSs based on the maximum received power \cite{yogi}.

\subsection{Channel Model}
Assuming Time Division Duplexing (TDD), the signal-to-interference-plus-noise ratio (SINR) of user $i$ from BS $b$ on subchannel $m$ is given as follows.
\begin{equation} \label{eq1}
    \gamma_i^{b,m}=\frac{P^{b,m} g_i^{b,m}}{\sum\limits_{\substack {\hat{b} \neq b \\ \hat{b} \in \Bb}}P^{\hat{b},m} g_{i}^{\hat{b},m} + \sigma^2 } \, ,
\end{equation}
where $P^{b,m} = \frac{P^{b}}{\text{M}}$ is the power allocated by the BS $b$ to the subchannel $m \in \Mb$, $P^{b}$ is the maximum transmit power of BS $b$, M is the total number of subchannels, $g_i^{b,m}=\|h_i^{b,m}\|^2$ is the channel gain between user $i$ and BS $b$ on subchannel $m$, ${\sum\limits_{\substack {\hat{b} \neq b \\ \hat{b} \in \Bb}}P^{\hat{b},m} g_{i}^{\hat{b},m}}$ is the interference on the subchannel $m$, and $\sigma^2$ is the noise power. The channel gain can be expressed as
\begin{equation} \label{eq2}
    g_i^{b,m}=10^{\frac{-pl(d_i^b)+g_t+g_r-f_s-v}{10}},
\end{equation}
where $pl(d_i^b)$ is the path loss between user $i$ and BS $b$ at a distance $d_i^b$, $g_t$ is the transmitter gain, $g_r$ is the receiver gain, $v$ is the penetration loss, and $f_s$ is the loss due to shadowing and fading. The corresponding link rate $r_i^{b,m}$ for a given SINR $\gamma_i^{b,m}$ in (\ref{eq1}) between user $i$ and BS $b$, can be expressed as
\begin{equation} \label{eq3}
   r_i^{b}=\frac{\eta(\gamma_i^{b,m}) sc_o sy_o}{t_{sc}} \text{M},
\end{equation}
where $\eta(\gamma_i^{b,m})$ can be obtained using the Adaptive Modulation and Coding Scheme (MCS) as given in \cite{yogi}. Further, $sc_o$, $sy_o$, and $t_{sc}$ represent the number of subcarriers per subchannel, the number of symbols per subchannel, and subframe duration (in seconds), respectively.

\subsection{CoMP}
We consider a set of CoMP clusters denoted by $\mathcal{C}=\lbrace 1,2,...,C \rbrace$ in a given area $\Ab$. The BSs are clustered using \textit{K}-means clustering approach. Let $\mathcal{I}_c$ and $\mathcal{I}_{nc}$ be the set of CoMP users and non-CoMP users, respectively, in cluster $c$ as shown in Fig. \ref{fig:system_model}. If $\gamma_{i}^m < \gamma_{th}$, then $i \in \Ib_c$, else the user belongs to $\Ib_{nc}$. Let the set of BSs in the CoMP cluster $c$ be denoted by $\mathcal{B}_c$, $\mathcal{B}_c=\lbrace 1,2,...,B_c \rbrace$. All the CoMP users in a cluster $c$ jointly receive information from the BSs for a duration of $\theta_c$, whereas, for the remaining $\left(1-\theta_c\right)$ time fraction, individual BSs in a cluster $c$ serve their respective non-CoMP users separately. Assuming each cluster $c$ has its typical value of $\theta_c$ \cite{yogi}, the SINR for the CoMP user $i$ in cluster $c$ is given as follows.
\begin{equation} \label{eq4}
    \gamma_{i}^m=\dfrac{\sum\limits_{\substack{l \in \Bb_{c}}}P^{l,m}g_{i}^{l,m}}{\sum\limits_{\substack{\hat{l} \in \Bb \\ \hat{l} \not\in \Bb_{c}}}P^{\hat{l},m}g_{i}^{\hat{l},m} + \sigma^{2}}, \forall i \in \Ib_c \,
\end{equation}
where $\sum\limits_{\substack{l \in \Bb_{c}}}P^{l,m}g_{i}^{l,m}$ is the power received by the user $i$ from all the BSs present in the cluster $c$, ${\sum\limits_{\substack{\hat{l} \in \Bb \\ \hat{l} \not\in \Bb_{c}}}P^{\hat{l},m}g_{i}^{\hat{l},m}}$ is the interference from other BSs in the system that do not belong to the CoMP cluster $c$, and $\sigma^2$ is the noise power. Similarly, the SINR of the non-CoMP users in cluster $c$ is computed using (\ref{eq1}).

\subsection{NOMA}
We now consider NOMA pairing for the CoMP based system. Let $\gamma_s$ and $\gamma_w$ be the OMA SINRs of strong and weak users in a NOMA pair, respectively, computed using (\ref{eq1}). We consider the minimum SINR difference (MSD) criteria given in \cite{mouni} to pair two users and calculate the optimal power fraction of the strong user. If the two users do not satisfy the MSD, then they are served as OMA users. The SINRs of paired NOMA users with perfect SIC are expressed as follows.
\begin{equation} \label{eq7a}
    \hat{\gamma}_s^{b,m}=\frac{\zeta_s P^{b,m} g_s^{b,m}}{\sum\limits_{\substack{\hat{b} \in \Bb \backslash b}}P^{\hat{b},m}g_s^{\hat{b},m} + \sigma^{2}},
\end{equation}
\begin{equation} \label{eq7b}
    \hat{\gamma}_w^{b,m}=\frac{(1-\zeta_s) P^{b,m} g_w^{b,m}}{\zeta_s P^{b,m}g_w^{b,m}+\sum\limits_{\substack{\hat{b} \in \Bb \backslash b}}P^{\hat{b},m}g_w^{\hat{b},m} + \sigma^{2}},
\end{equation}
where $\zeta_s$ is the optimal power fraction allocated to the strong user in the pair computed as in \cite{mouni}, $\hat{\gamma}_s^{b,m}$ is the SINR of the strong user with perfect SIC after NOMA pairing, $P^{b,m}$ is the total power assigned to the pair, $g_s^{b,m}$ is the channel gain of the strong user as in (\ref{eq2}), $\sum\limits_{\substack{\hat{b} \in \Bb \backslash b}}P^{\hat{b},m}g_s^{\hat{b},m}$ is the aggregate interference received from other BSs, $\hat{\gamma}_w^{b,m}$ is the SINR of the weak user after NOMA pairing, $(1-\zeta_s)$ is the power fraction allocated to the weak user, and $g_w^{b,m}$ is the channel gain of the weak user as in (\ref{eq2}). Throughout the work, we consider the adaptive user pairing algorithm (AUP) proposed in \cite{mouni} which uses MSD criteria for pairing users in two groups. However, the formation of the strong and weak user groups for a CoMP--NOMA-based system is not obvious. To address this issue, we next propose the grouping and pairing scheme.
\section{Proposed CoMP--NOMA user pairing scheme} \label{pairing}
There are three types of NOMA pairs possible based on the users present in a cluster: CoMP--CoMP, (non-CoMP)--CoMP, and (non-CoMP)--(non-CoMP). In this work, due to the given space constraints, we study the performance of the CoMP--NOMA system by proposing a user grouping and pairing scheme considering only CoMP--CoMP and (non-CoMP)--(non-CoMP) NOMA pairs. 
\begin{figure}[t]
     \centering
     \includegraphics[width=9cm,height=14cm,keepaspectratio]{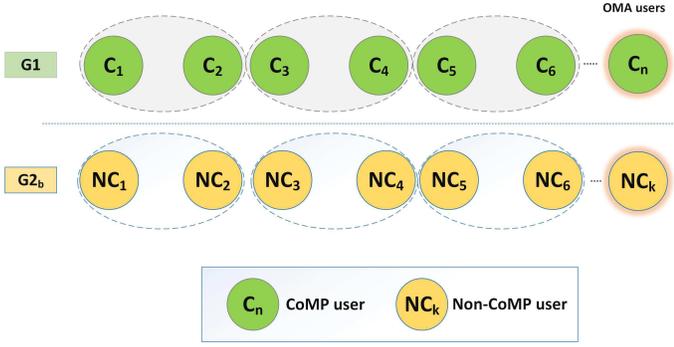}
     \caption{Illustration of the proposed grouping and pairing of users.}\vspace{-0.18in}
     \label{fig:scheme_2}
\end{figure}

The group of all CoMP users is designated as group $\textbf{G1}$ and the group of non-CoMP users associated with individual base stations is denoted as group $\textbf{G2}_{b}$. After obtaining the groups, NOMA pairing is done as per the AUP mentioned in \cite{mouni}. However, in this scheme NOMA pairing is done within the group as shown in Fig. \ref{fig:scheme_2}. Therefore, no cross-group pairing occurs. The SINR of a weak CoMP user in the CoMP--CoMP pair according to this scheme is as follows.
\begin{equation} \label{c1}
    \Bar{\gamma}_{i_w}^m= \frac{\sum\limits_{\substack{t \in \Bb_{c}}} (1-\zeta_t) P^{b,m} g_{i_w}^{t,m}}{\sum\limits_{\substack{t \in \Bb_{c}}} \zeta_k P^{b,m}g_{i_w}^{t,m} + \sum\limits_{\substack{\hat{l} \in \Bb \\ \hat{l} \not\in \Bb_{c}}}P^{\hat{l},m}g_{i_w}^{\hat{l},m} + \sigma^{2}},
\end{equation}
where $\Bar{\gamma}_{i_w}^m$ is the SINR of the weak user in the CoMP--CoMP pair in cluster $c$ and $(1-\zeta_t)$ is the power fraction allocated by each BS in cluster $c$ to the weak CoMP user. The SINR of a strong user with perfect SIC in the CoMP--CoMP pair in cluster $c$ is given as follows.
\begin{equation} \label{c2}
    \Bar{\gamma}_{i_s}^m= \frac{\sum\limits_{\substack{t \in \Bb_{c}}}\zeta_t P^{b,m} g_{i_s}^{t,m}}{\sum\limits_{\substack{\hat{l} \in \Bb \\ \hat{l} \not\in \Bb_{c}}}P^{\hat{l},m}g_{i_s}^{\hat{l},m} + \sigma^{2}},
\end{equation}
where $\Bar{\gamma}_{i_s}^m$ is the SINR of the strong user with perfect SIC in the CoMP--CoMP pair and $\zeta_t$ is the power fraction allocated to the strong CoMP user by each BS in the cluster $c$. The SINRs of strong and weak users of (non-CoMP)--(non-CoMP) pair are computed as given in (\ref{eq7a}) and (\ref{eq7b}), respectively. The CoMP pairs are served collectively by all the BSs in the cluster during the duration of $\Bar{\theta}_c$. Using an $\alpha$-fair scheduler ($\alpha$ is the fairness parameter \cite{yogi}), each CoMP--CoMP NOMA pair or OMA CoMP user (if any) is given a time fraction of $\Bar{\beta}_{i}$ which is given as follows (for $\alpha=1$)\cite{yogi}.
\begin{equation}
    \Bar{\beta}_{i} = \frac{1}{|\Hat{\Ib}_c| + |\Tilde{\Ib}_c|},
\end{equation}
where $|X|$ denotes the cardinality of set $X$, $\Hat{\Ib}_c$ represents the set of CoMP--CoMP NOMA pairs formed, and $\Tilde{\Ib}_c$ is the set of OMA CoMP users which could not be paired (if any). The scheduling time fraction for (non-CoMP)--(non-CoMP) NOMA pairs that are served by their respective BSs in the duration of $(1-\Bar{\theta}_c)$ with an $\alpha$-fair scheduler is given as follows (for $\alpha=1$)\cite{yogi}.
\begin{equation}
    \Bar{\beta}_{i}^{b} = \frac{1}{|\Hat{\Ib}_{nc}^{b}| + |\Tilde{\Ib}_{nc}^{b}|},
\end{equation}
where $\Hat{\Ib}_{nc}^{b}$ represents the set of (non-CoMP)--(non-CoMP) NOMA pairs formed for a BS $b$ in the cluster $c$, $\Tilde{\Ib}_{nc}^{b}$ is the set of OMA non-CoMP users for a BS $b$ in the cluster $c$ which could not be paired (if any). We consider the expressions of optimal $\Bar{\theta}_c$, $\Bar{\beta}_{i}$, and $\Bar{\beta}_{i}^{b}$ computed for a purely CoMP based system in \cite{yogi} to study the performance of this scheme. However, they may be sub-optimal for the current CoMP--NOMA-based system.

\begin{table}[t]
\caption{Simulation Setup}
\begin{center}
\begin{tabular}{| m{4.4cm} | m{3cm}|}
\hline
\textbf{\textit{Parameter}}& \textbf{\textit{Value}} \\
\hline
Area, $\Ab$ ($\text{km}^2$) & 25 \\
\hline
AWGN Power spectral density (dBm) & $-174$. \\
\hline
Base Station density, $\lambda_{b}$ ($/\text{km}^2$) & 16, 30\\
\hline
Fairness parameter, $\alpha$ & 1\\
\hline
Number of subcarriers per subchannel, $sc_o$ & $12$ \\
\hline
Number of symbols per subcarrier, $sy_o$ & $14$ \\
\hline
Number of clusters, $K$ & $10$ \\
\hline
Number of iterations & $10^3$ \\
\hline
Path Loss ($d$ is in kms) & $133.6 + 35\log_{10}(d) + \chi$ \\
\hline
Standard deviation of shadowing random variable (dB) & 8\\
\hline
Subchannel Bandwidth (kHz) & $180$ \\
\hline
Total number of subchannels, M & $100$ \\
\hline
Transmission power, $P^{b}$ (dBm) & $46$ \\
\hline
User density, $\lambda_u$ ($/\text{km}^2$) & $40,50,60,70,80,90,140,\newline150$\\
\hline
\end{tabular}\vspace{-0.2in}
\label{table}
\end{center}
\end{table}
\section{Results and Discussions} \label{results}
In this section, we present the numerical results to evaluate the performance of the proposed scheme through Monte Carlo simulations performed in MATLAB for different combinations of $\lambda_b$, $\lambda_u$, and $\gamma_{th}$. The details of the simulation parameters are summarized in Table \ref{table}.
\begin{figure}[t]
    \centering
    \includegraphics[width=9cm,height=10cm,keepaspectratio]{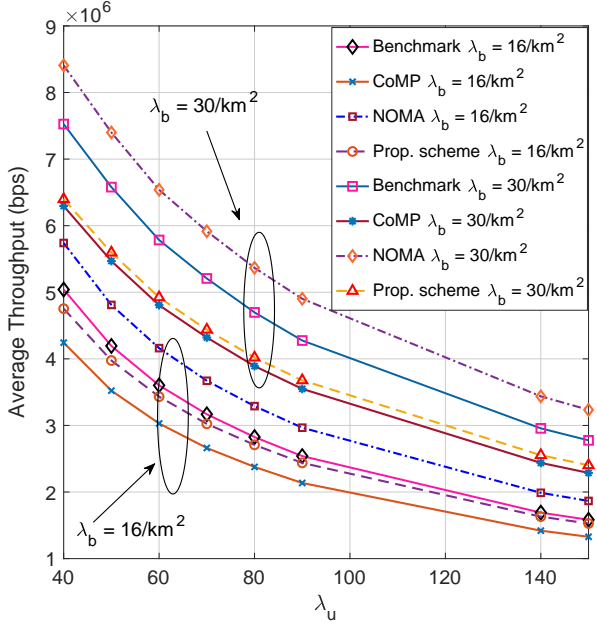}\vspace{-0.1in}
    \caption{Illustration of variation of throughput with user density for $\gamma_{th} = -6.5$ dB and $\lambda_{b} = 16/\text{km}^2$ and $\lambda_{b} = 30/\text{km}^2$ .}\vspace{-0.19in}
    \label{fig:throughput_1}
\end{figure}
Fig. \ref{fig:throughput_1} shows the variation of average throughput for a $\gamma_{th}$ of -6.5 dB with respect to $\lambda_u$ for $\lambda_b=16 /\text{km}^2$ and $30 /\text{km}^2$, respectively. It is observed from Fig. \ref{fig:throughput_1} that with an increase in user density there is a decrease in average throughput of the proposed scheme. The proposed scheme is offering significantly better throughput than that of the CoMP only system for lower values of $\lambda_u$. However, the performance of this scheme is almost equal to that of CoMP at higher $\lambda_u$. Similar trends are observed for higher values of $\lambda_{b}$, but for all values of $\lambda_{u}$. The good performance of the proposed scheme for lower $\lambda_u$ in the case of $\lambda_b = 16/\text{km}^2$ can be attributed to the availability of more time fraction to the users at lower $\lambda_u$. As $\lambda_u$ increases, $\Bar{\theta}_c$ increases due to the increase in the number of CoMP--CoMP NOMA pairs and this degrades the performance of the proposed scheme. At significantly higher $\lambda_{b}$, a few NOMA pairs are formed and hence the performance of the proposed scheme is similar to that of the CoMP only system.
\begin{figure}[t]
     \centering
     \includegraphics[width=9.5cm,height=12.5cm,keepaspectratio]{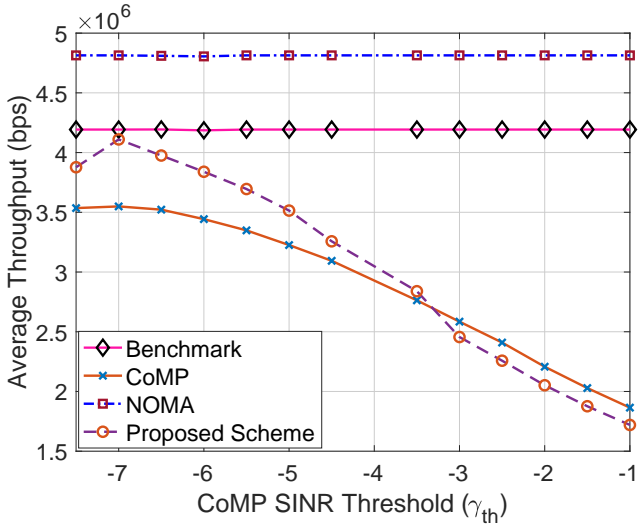}\vspace{-0.06in}
     \caption{Illustration of variation of throughput with $\gamma_{th}$ for $\lambda_{b} = 16/\text{km}^2$ and $\lambda_{u}=50/\text{km}^2$.}\vspace{-0.19in}
     \label{fig:throughput_3}
\end{figure}

The variation of average throughput with respect to $\gamma_{th}$ for Benchmark, CoMP only, NOMA only, and the proposed scheme is shown in Fig. \ref{fig:throughput_3}. It can be observed from the plot that the proposed scheme's performance is better than that of the CoMP only system for lower values of $\gamma_{th}$. After a certain $\gamma_{th}$, the CoMP only system is performing better than the proposed scheme due to the increase in CoMP users in a cluster because of which the time fraction available for non-CoMP NOMA pairs reduces.

\begin{figure}[t]
     \centering
     \includegraphics[width=9.1cm,height=12.1cm,keepaspectratio]{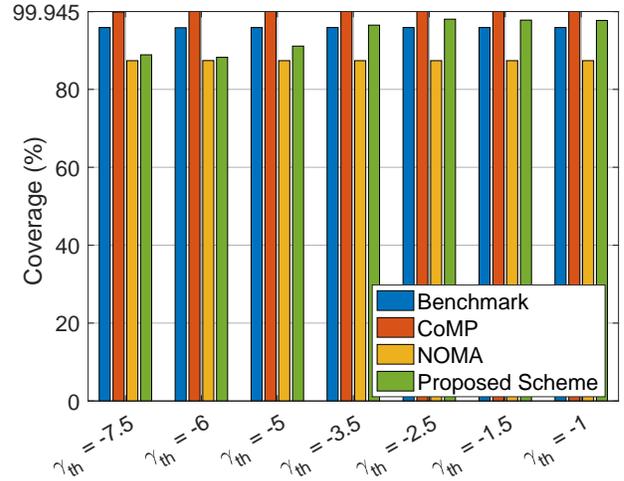}\vspace{-0.1in}
     \caption{Illustration of variation of coverage of different schemes with $\gamma_{th}$ for $\lambda_{b} = 16/\text{km}^2$.}\vspace{-0.17in}
     \label{fig:coverage_1}
\end{figure}

Similarly, the variation of coverage with respect to $\gamma_{th}$ is shown in Fig. \ref{fig:coverage_1}. It is observed that coverage increases with an increase in $\gamma_{th}$. The coverage of the CoMP only system is the highest in comparison to the benchmark followed by the proposed scheme, and NOMA only systems. CoMP only system performs the best in terms of Coverage at the cost of reduced throughput as dedicated CoMP resources are available for users with lower channel gains. Thus, CoMP--NOMA system with the proposed scheme performs better than the traditional CoMP in terms of throughput and traditional NOMA in terms of coverage.

\section{Conclusion} \label{conclusion}
In this work, we have studied the performance of the CoMP--NOMA system by proposing a user pairing scheme considering only two types of pairs. The proposed scheme results in enhanced coverage as compared to a NOMA-based system. However, this enhancement comes at a cost of reduced network throughput although, superior throughput trends are observed when compared to CoMP only system for certain $\gamma_{th}$, $\lambda_u$, and $\lambda_b$. Thus, the proposed scheme offers a trade-off between a CoMP system at one extreme with high coverage-low network throughput and a NOMA system at the other extreme with low coverage-high throughput. The presented scheme can be used by cellular network planners to appropriately select the coverage-network throughput trade-off by dynamically tuning the threshold and other parameters. In the future, we plan to study this dynamic tuning using state-of-the-art machine learning algorithms.
\section{Acknowledgement}
This work was supported in part by the Indo-Norway projects, no. 287918, 280835, and DST NMICPS through TiHAN Faculty fellowship of Dr. Abhinav Kumar.

\end{document}